\documentclass[showpacs,amsmath,amssymb,aps,twocolumn,longbibliography,pra]{revtex4-2}%,superscriptaddress
\usepackage{graphicx}
\usepackage[colorlinks=true,citecolor=blue,linkcolor=magenta]{hyperref}
\usepackage[usenames]{color}
\usepackage{relsize}
\usepackage[english]{babel}
\usepackage{enumerate}
\usepackage{bbm}
\usepackage{xcolor}
\usepackage{framed}
\usepackage{bm}

\newcommand{\ket}[1]{\left| #1 \right>} % for Dirac bras
\newcommand{\bra}[1]{\left< #1 \right|} % for Dirac kets
\newcommand {\grsim} {\ {\raise-.5ex\hbox{$\buildrel>\over\sim$}}\ }
\newcommand {\lessim} {\ {\raise-.5ex\hbox{$\buildrel<\over\sim$}}\ }

\newcommand{\create}[1]{\hat{#1}^{\dagger}}

\usepackage{xcolor}
\usepackage{siunitx}
\usepackage{booktabs}% http://ctan.org/pkg/booktabs

\newcommand{\nocontentsline}[3]{}
\newcommand{\tocless}[2]{\bgroup\let\addcontentsline=\nocontentsline#1{#2}\egroup}

\newcommand{\RN}[1]{%
  \textup{\uppercase\expandafter{\romannumeral#1}}%
}

\usepackage{xr}
\begin{document}
%TC:ignore
%\title{Error characterization in quantum simulation of many-body physics}
%\title{Optimal atomic entangled states for quantum sensing with lossy state preparation}
\title{Optimizing lossy state preparation for quantum sensing using Hamiltonian engineering }
%\title{Benchmarking quantum simulation of many-body physics using operator thermalization}
\author{Bharath Hebbe Madhusudhana$^{1}$ }
%\author{Fermi1-team$^{1,2,3}$}
%\author{Bharath~Hebbe~Madhusudhana$^{1,2,3}$, Sebastian~Scherg$^{*1,2,3}$, Thomas~Kohlert$^{*1,2,3}$, Immanuel~Bloch$^{1,2,3}$,  Monika~Aidelsburger$^{1,2}$}

\affiliation{$^{1}$MPA-Quantum, Los Alamos National Laboratory, Los Alamos, NM-87544, United States}

%\pacs{XXX}
%TC:endignore
%TC:break abstract
%about 7 summarising lines

\begin{abstract} 
One of the most prominent platforms for demonstrating quantum sensing below the standard quantum limit is the spinor Bose-Einstein condensate.While a quantum advantage using several tens of thousands of atoms has been demonstrated in this platform, it faces an important challenge: atom loss. Atom loss is a Markovian error process modelled by Lindblad jump operators, and a no-go theorem, which we also show here, states that the loss of atoms in all spin components reduces the quantum advantage to a constant factor. Here, we show that this no-go theorem can be circumvented if we constrain atom losses to a \textit{single} spin component. Moreover, we show that in this case, the maximum quantum Fisher information with $N$ atoms scales as $N^{3/2}$,  establishing that a \textit{scalable} quantum advantage can be achieved \textit{despite} atom loss. Although Lindblad jump operators are generally non-Hermitian and non-invertible, we use their \textit{Moore-Penrose inverse} to develop a framework for constructing several states with this scaling of Fisher information in the presence of losses. We use Hamiltonian engineering with realistic Hamiltonians to develop experimental protocols for preparing these states. Finally, we discuss possible experimental techniques to constrain the losses to a single spin mode.
\end{abstract}

\maketitle

\section{Introduction}
\begin{figure}[ht!]
\includegraphics[scale=0.34]{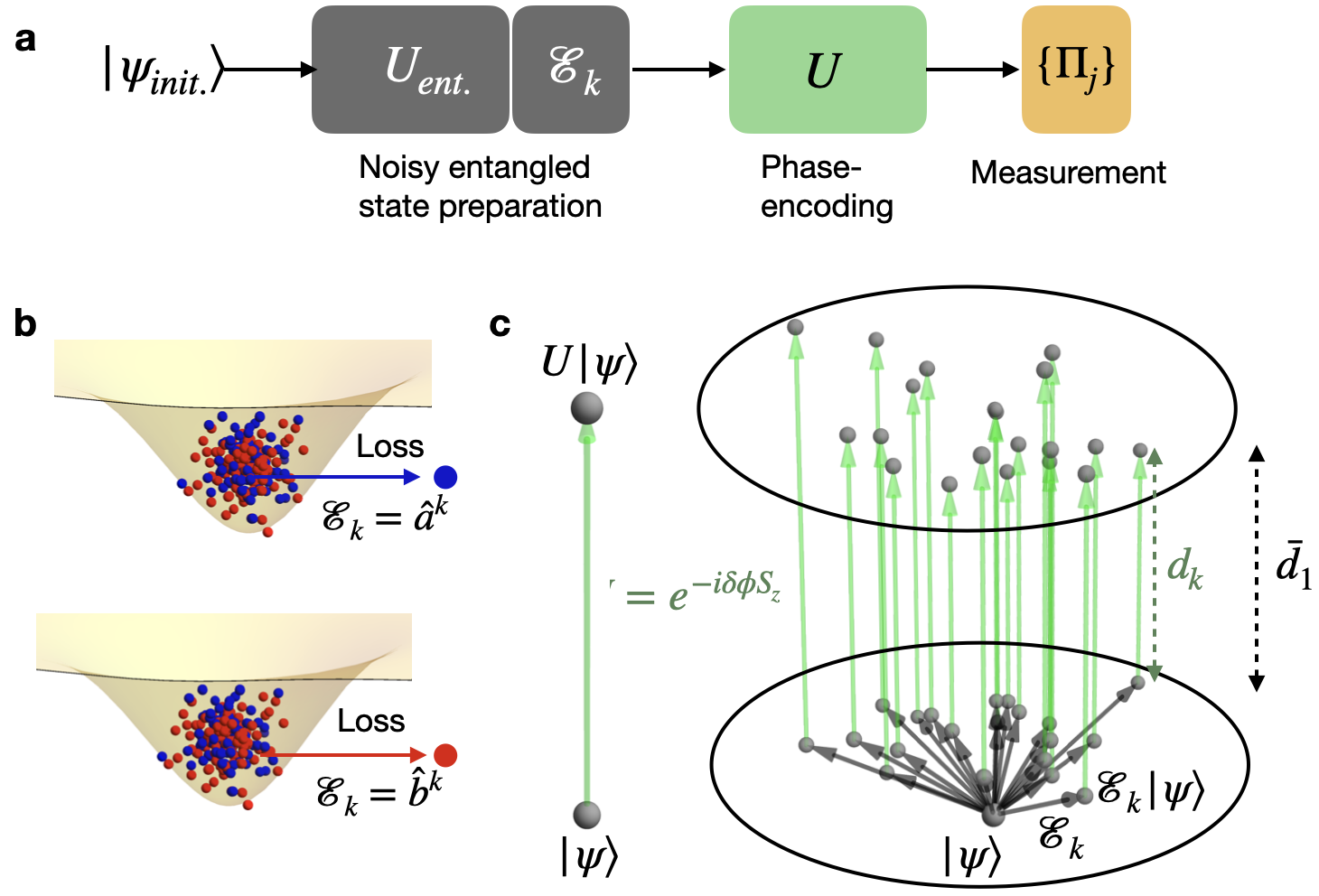}
\caption{\textbf{Losses and Quantum Fisher information:} \textbf{a.}  schematic of the quantum sensing sequence, with a noise process assumed to occur during state preparation.  It is modelled by a random noise/loss operator $\mathcal E_k$ acting on the state prepared.  \textbf{b.} A two mode bosonic system is typically affected by two common types of losses --- loss of atoms in the two modes ($\create{a}\ket{\text{vac}}$ and $\create{b}\ket{\text{vac}}$, see text).  \textbf{c.} (left) the unitary $U=e^{-i\delta \phi S_z}$ generated by the Hamiltonian $S_z$ \textit{moves} a state within the Hilbert space from $\ket{\psi}$ to $U\ket{\psi}$.  The distance between them is a measure of the sensitivity, quantified by the quantum Fisher information. (right) in the presence of probabilistic errors $\mathcal E_k$,  the state is moved randomly to $\mathcal E_k \ket{\psi}$, which under $U$ moves by a distance of $d_k$. There are two ways to define the sensitivity --- distance between the average of the states before and after the application of $U$ and average of the distances $d_k$. They correspond to quantum Fisher informations $\mathcal F^{(1)}_Q$ and $\mathcal F^{(2)}_Q$ respectively (see text).   }\label{Fig1}
\end{figure}

Practical quantum advantage in most quantum technologies is limited by noise, decoherence and losses in quantum control.  Consequently, the study of the physical origins of the noise and techniques to mitigate them or correct them has emerged as a dominant research area over the last few years.  In particular, in quantum sensing, while a quantum advantage has been demonstrated in several different experimental platforms including cold atoms~\cite{Pezze:2018},  NV-centers~\cite{Aslam2023} and photons~\cite{Pirandola2018},  its scalability is limited by noise and losses in quantum control.  Fig.~\ref{Fig1}a shows a schematic of a quantum sensing protocol with three steps: state preparation, phase encoding where the unknown phase which we intend to measure is encoded into the system, and measurement.  The sensitivity is quantified by the quantum Fisher information $\mathcal F_Q$ of the state prepared in the first step, which scales as $N$(the standard quantum limit (SQL)) for $N$ unentangled atoms (or photons) and up to $N^2$ (the Heisenberg limit (HL)) for certain entangled states and therefore, the quantum advantage is quantified by $\mathcal F_Q/N$.  

A central result in noisy quantum sensing is the identification of a class of Markovian noise models, under which the optimal quantum Fisher information is back to the SQL scaling,  leaving at-most a constant factor quantum advantage.  This class of noise models preclude any significant quantum advantage. They include the so-called \textit{phase covariant} and \textit{time-homogeneous}  processes~\cite{PhysRevLett.79.3865, Demkowicz_Dobrza_ski_2012, Escher_2011, PhysRevA.83.021804}. That is, the noise operators that commute with the phase-encoding Hamiltonian and are homogeneous in time.   

Noise models outside of this class are known to allow a scalable quantum advantage in the quantum Fisher information. For instance,  certain phase covariant but time- inhomogeneous noise models have been shown to follow a typical scaling of $\mathcal F_Q \sim N^{3/2}$~\cite{PhysRevA.84.012103, PhysRevLett.109.233601, PhysRevLett.116.120801}, which came to be known as the \textit{zeno regime}.  Another set of examples are noise models that are not phase covariant. A scaling of $\mathcal F_Q \sim N^{5/3}$ have been established for noise models that represent a dephasing in a direction normal to the phase-encoding unitary~\cite{PhysRevLett.111.120401} and therefore are not phase covariant.  Moreover, some noise models that are neither phase-covariant, nor time-homogeneous are known to feature a scaling of $\mathcal F_Q\sim N^{7/4}$~\cite{Haase_2018} in a system of qubits coupled to a bosonic bath.

Considering these results, a promising approach to obtain scalable, noisy quantum advantage is to \textit{engineer} the noise.  For instance,  minimizing the phase covariant part of the noise, at the cost of an increased  non-phase-covariant part of the noise can improve the scaling of the quantum Fisher information.  Identifying noise models that allow for a scalable quantum advantage is therefore important from an experimental perspective~\cite{PhysRevX.5.031010}.

Another approach is to use quantum error correction in sensing. While quantum error correction applied to computation is a vibrant and growing area, its applications to quantum sensing has received sparse attention so far. One of the central results in this area is the ``Hamiltonian not in Lindblad Space (HNLS)" criterion~\cite{PRXQuantum.2.010343, PhysRevX.7.041009, Sekatski2017quantummetrology}. The latter is a necessary and sufficient criterion for the existence of quantum error correction codes that would recover the Heisenberg limit in the presence of Markovian noise. However, the practical application of this criterion requires noiseless ancilla qubits. 

Other results~\cite{PhysRevLett.122.040502} have shown that error correction protocols that don't use ancilla qubits exist, albeit for a special class of errors. More recently, the metrological performance of more realistic protocols was analyzed~\cite{PhysRevLett.133.170801}, showing that a dephasing error precludes a quantum advantage in almost all of these protocols. On the experimental side, an error correction technique called erasure conversion, which has been used very successfully in quantum computation was applied to quantum sensing recently~\cite{PhysRevLett.133.080801}. 

Most of the results so far have been focussed on qubit systems, i.e., a set of $N-$qubits with a noise channel acting independently on each qubit.  Moreover, the class of noise channels that allow for a scalable quantum advantage hasn't been fully explored.  Here, we consider a system of $N$ identical two mode bosons. While this is a very popular platform for quantum sensing both experimentally~\cite{Gross_2010, Hamley_2012, PhysRevLett.113.103004, PhysRevLett.109.253605, PhysRevLett.116.093602, Eckner_2023} and theoretically, the problem of noisy sensing in this platform has received sparse attention~\cite{PhysRevA.82.053804}.  One of the most prominent sources of Markovian noise in this system is loss of atoms. This includes two independent noise channels, one corresponding to each mode.  We show that the presence of losses in both modes limits the quantum advantage to a constant factor~\cite{PhysRevA.82.053804}.We show that when the losses are dominant in only one of the modes, one can surpass the SQL, with a scaling $\mathcal F_Q\sim N^{3/2}$, independent of rate of loss. We also develop a technique based on the idea of  \textit{Moore-Penrose} inverse to construct quantum states that feature this scaling. We use Hamiltonian engineering~\cite{PRXQuantum.4.040333} to develop protocols to prepare these states using experimentally implementable Hamiltonians. Finally, we discuss experimental techniques to confine loss in one mode.

\section{Problem set-up and figures of merit}
We consider a system of $N$ identical bosons with two internal states, $\hat{a}^{\dagger}\ket{\text{vac}}$ and $\hat{b}^{\dagger}\ket{\text{vac}}$. Here $\create{a}$ and $\create{b}$ are the creation operators corresponding to the two modes and $\ket{\text{vac}}$ is the vacuum state.  Examples of such a system include two component BECs, which has been extensively used to demonstrate entangled quantum sensing protocols and photonic systems with two modes. States of this system belong to an $N+1$ dimensional Hilbert space. A state $\ket{\psi}$ can be written as $\ket{\psi}= \sum_{n=0}^N \frac{C_n}{\sqrt{n!(N-n)!}} (\create{a})^n(\create{b})^{N-n}\ket{\text{vac}}$. For instance, the well known coherent state is represented as $\ket{\psi_{\text{cohr}}}= \frac{1}{2^{N/2}}\left(\create{a}+\create{b}\right)^N\ket{\text{vac}}= \frac{1}{2^{N/2}} \sum_n \binom{N}{n}(\create{a})^n(\create{b})^{N-n}\ket{\text{vac}}$.  The Greenberger-Horne-Zeilinger (GHZ) state would be $\ket{\psi_{\text{GHZ}}}=\frac{1}{\sqrt{2N!}}\left((\create{a})^N +(\create{b})^N\right)\ket{\text{vac}}$.

The two internal states can be mapped to a spin-1/2 system and the Hilbert space can be mapped to the symmetric subspace of $N$ spin-1/2 systems. We consider quantum sensing protocols to measure an unknown phase $\phi$ imparted by a unitary $U_{\phi}=e^{-i \phi S_z}$ where, the operator $S_z$ is defined as
\begin{equation}
S_z = \frac{1}{2}(\create{a}\hat{a}-\create{b}\hat{b})
\end{equation}

A standard interferometry in this system involved preparation of the atoms in the state $\ket{\psi}$, evolution under $U_{\phi}$, $\ket{\psi}\mapsto e^{-i \phi S_z}\ket{\psi}$, followed by a measurement in an appropriate basis to estimate $\phi$ (Fig.~\ref{Fig1}a) .  A figure-of-merit, independent of the basis of measurement is given by the quantum Fisher information, $\mathcal F_Q(\ket{\psi}, S_z)$, which is a measure of the  sensitivity of the state $\ket{\psi}$ to evolution under $S_z$. We assume that the measurement is repeated $\nu$ times. The precision of the estimate of $\phi$ is related to the quantum Fisher information as $\Delta \phi \geq \frac{1}{\sqrt{\nu \mathcal F_Q(\ket{\psi}, S_z)}}$.

%\begin{figure*}[ht!]
%\includegraphics[scale=1]{Figure_2.pdf}
%\caption{\textbf{Lossy fisher information:} \textbf{a.} Fisher information $\mathcal F^{(2)}$ as a function of the loss probability for $L=\hat{a}$ (the curves are similar for $L=\hat{b}$) for three states: (i) OAT, (ii) TACT and (iii) GHZ (see text). $N=100$ for all three cases. \textbf{b.} $\mathcal F^{(2)}$ as a function of $N$ for $L=\hat{a}$, for the same states. The loss probability $p=0.1$.  \textbf{c, d.} similar plots for $L=\create{a}\hat{b}$. }\label{Fig2}
%\end{figure*}

 \begin{figure*}[ht!]
\includegraphics[scale=0.55]{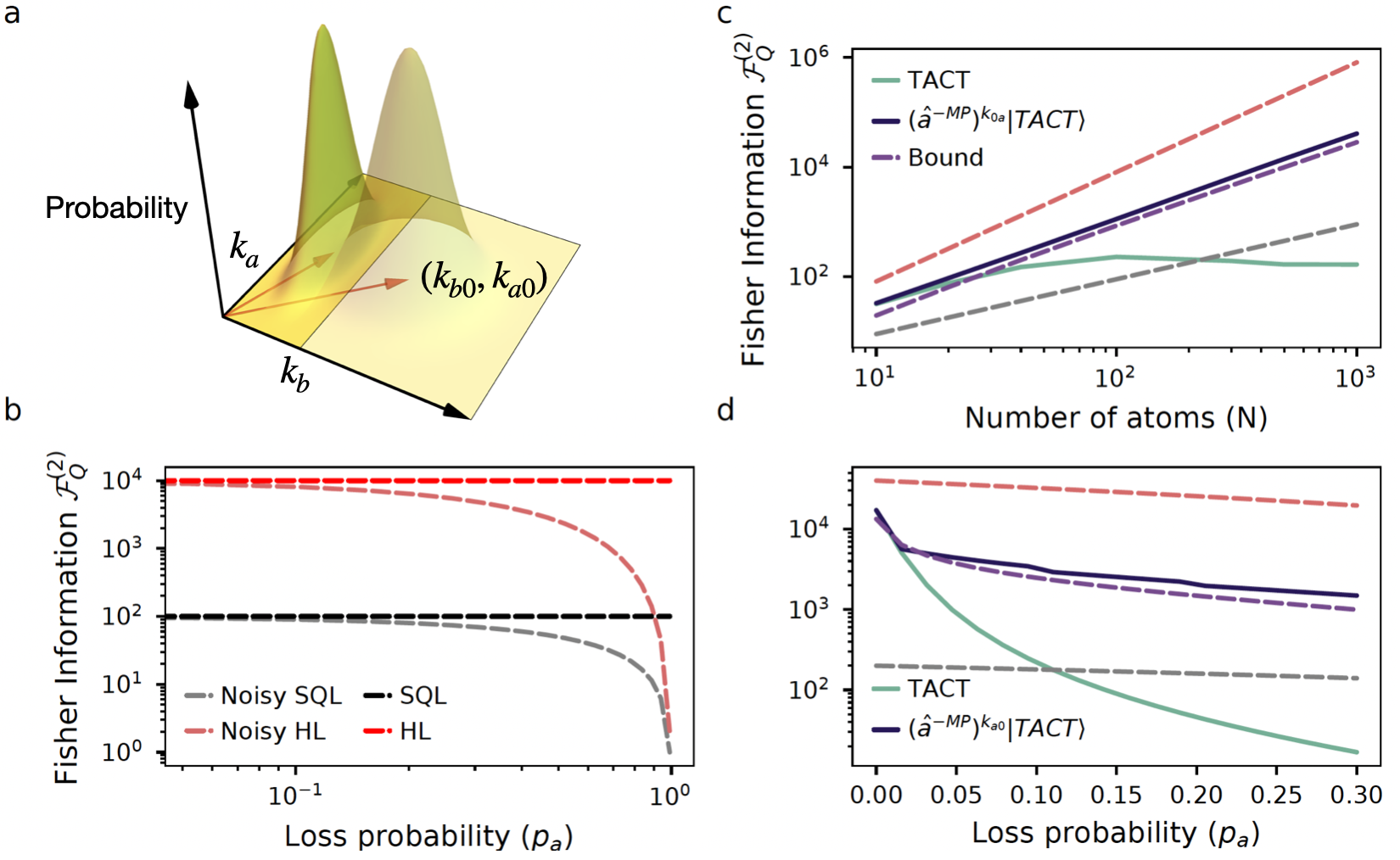}
\caption{\textbf{Noisy Fisher information and Moore-Penrose inversion}: \textbf{a.} Probability distribution $\binom{N}{k_a}\binom{N}{k_b}p_a^{k_a}(1-p_a)^{N-k_a}p_b^{k_b}(1-p_b)^{N-k_b}$ of error with two losses, $\hat{a}$ and $\hat{b}$. The yellow surface represents the case $p_a\approx p_b$ and the greenish surface represents the case $k_b\approx 0$, i.e., single loss channel. \textbf{b.} Noisy SQL and noisy HL (Eq.~(\ref{NoisyHL})) as a function of $p$. \textbf{c.} $\mathcal F^{(2)}_Q$ for the Moore-Penrose inverse, $(\hat{a}^{-MP})^{k_{a0}}\ket{{TACT}}$ of the two-axis counter twisted state, with $k_{a0} = \lfloor (1-p_a)N\rfloor$ as a function of $N$ for $p_a=0.1$. It maybe compared with $\mathcal F^{(2)}_Q$ without the Moore-Penrose inverse.  The purple dashed curve is the bound Eq.~(\ref{lower_bound}) with the $N^{3/2}$ scaling. \textbf{d.} similar plots, as a function of $p_a$ and $N=200$.  Note that in both \textbf{c} and \textbf{d}, the Moore-Penrose inverted state is well above the SQL whereas the bare TACT state quickly goes below. }\label{Fig3}
\end{figure*}

The quantum Fisher information can be interpreted as the Fubini-Study distance travelled by the state $\ket{\psi}$ during an evolution under $S_z$. That is, the distance between the state $\ket{\psi}$ and $e^{-i \phi S_z}\ket{\psi}$, given by 
\begin{equation}
d(\ket{\psi}, e^{-i \phi S_z}\ket{\psi}) = \text{arccos}\left(|\bra{\psi}e^{-i \phi S_z}\ket{\psi}\rangle|\right)
\end{equation}
Under for an infinitesimal $\delta \phi$, the distance expands as:
\begin{equation}
\begin{split}
d(\ket{\psi}, e^{-i \delta \phi S_z}\ket{\psi}) &=\\
 \text{arccos}&\left(|1-i \delta \phi\bra{\psi}S_z\ket{\psi} -\frac{\delta \phi^2}{2}\bra{\psi}S_z^2\ket{\psi} |\right)\\
=\delta\phi^2 (&\bra{\psi}S_z^2\ket{\psi}-\bra{\psi}S_z\ket{\psi}^2)
\end{split}
\end{equation}
The objective of this work is to study the optimal states $\ket{\psi}$ for quantum sensing in the presence of standard Markovian noise. While the quantum Fisher information is a convenient figure of merit, it does not have a unique generalization in the presence of losses.  We model the Markovian losses which occur during the evolution under $S_z$, by the Gorini-Kosskowski-Lindblad-Sudarshan (GKLS) equation:
\begin{equation}\label{GKLS}
\begin{split}
\dot{\rho} =-i[S_z, \rho] - \sum_j \gamma_j\left( L_j\rho L_j^{\dagger} -\frac{1}{2}\left\{\rho, L_jL_j^{\dagger}\right\} \right)
\end{split}
\end{equation}
Here, $\rho=\ket{\psi}\bra{\psi}$ and $L_j$ are the Lindblad jump operators.  The first term in the R.H.S $-i[S_z, \rho]$ represents the unitary evolution under $S_z$.  Below, we will consider two types of losses corresponding to Lindblad operators: $L=\hat{a}$, representing loss of atoms in mode $a$, and  $L=\hat{b}$, representing loss of atoms in mode $b$.

There are two generalizations of the quantum Fisher information in the presence of Markovian losses.  For the following discussion, we will assume that there is only one Lindblad jump operator $L$ for conceptual simplicity.  There is a stochastic interpretation of Eq.~(\ref{GKLS}), where, the evolution over an infinitesimal $\delta \phi$ can be broken into two processes: (i)a stochastic jump $\ket{\psi}\mapsto L\ket{\psi}$, on \textit{each} of the $N$ atoms and (ii) a unitary evolution $e^{-i\delta \phi S_z}$. The former occurs  with a probability $p = \gamma dt$. Given the $N$ atoms, there are $N+1$ possible events, given by $L^k$, with $k=0, 1, \cdots, N$. The probabilities of these events is $p^k(1-p)^{N-k}\binom{N}{k}$. Therefore, the process moves the state $\ket{\psi}$ to:
\begin{equation}
\ket{\psi}\mapsto e^{-i\delta \phi S_z} \frac{L^k\ket{\psi}}{||L^k\ket{\psi}||}
\end{equation}
The solution to Eq.~(\ref{GKLS}) is approximated by the incoherent average of these states.  In our model, we assume $\nu$ experimental shots, where the i-th shot is modelled by the process $\ket{\psi}\mapsto e^{-i\delta \phi S_z} \frac{L^k_i\ket{\psi}}{||L^k_i\ket{\psi}||}$, where $k_i$ is a random integer sampled from the distribution $p^k(1-p)^{N-k}\binom{N}{k}$.  The average state would be $\rho(p)= \sum_{k=0}^N \binom{N}{k}p^k (1-p)^{N-k} \frac{L^k\ket{\psi}\bra{\psi}(L^{\dagger})^k}{||L^k\ket{\psi}||^2}$

Therefore, there are two ways to define the sensitivity to $S_z$: we may consider the \textit{averaged} distance between $e^{-i\delta \phi S_z} \frac{L^k\ket{\psi}}{||L^k\ket{\psi}||}$ and $ \frac{L^k\ket{\psi}}{||L^k\ket{\psi}||}$ or, the distance between the averaged state $\rho(p)$ and $e^{-i\delta \phi S_z}\rho(p)e^{i\delta \phi S_z}$.  We represent the latter by $\mathcal F^{(1)}_Q(\ket{\psi, S_z, L, p})$:
\begin{equation}
\begin{split}
\mathcal F^{(1)}_Q(\ket{\psi}, S_z, L, p) =\mathcal  F_Q \left(\rho(p), S_z\right)\\
\rho(p)= \sum_{k=0}^N \binom{N}{k}p^k (1-p)^{N-k} \frac{L^k\ket{\psi}\bra{\psi}(L^{\dagger})^k}{||L^k\ket{\psi}||^2}
\end{split}
\end{equation}

We represent the averaged sensitivity by $\mathcal F^{(2)}_Q(\ket{\psi}, S_z, L, p)$:
\begin{equation}\label{F2}
\begin{split}
\mathcal F^{(2)}_Q(\ket{\psi}, S_z, L, p) = \sum_{k=0}^N \binom{N}{k}p^k (1-&p)^{N-k}\\
\times &\mathcal  F_Q \left(\frac{L^k\ket{\psi}}{||L^k\ket{\psi}||}, S_z\right)
\end{split}
\end{equation}
Ref.~\cite{PhysRevA.91.042104} shows that $\mathcal F^{(2)}_Q\geq F^{(1)}_Q$.  Physically,  one can interpret them in the following way.  $\mathcal F^{(1)}_Q$ is the sensitivity if the integer $k_i$ in the dataset is \textit{unknown} and $\mathcal F^{(2)}_Q$ is the sensitivity if $k_i$ is \textit{known} for each $i=1, \cdots, \nu$ in the dataset. In the latter case, each one can effectively choose the optimal measurement to be performed for each shot, after knowing the loss process that occurred.  This is the intuitive reasoning behind the inequality $\mathcal F^{(2)}_Q\geq \mathcal F^{(1)}_Q$.  The knowledge of which loss process occurred can be obtained either using the number of atoms left at the end of the measurement, or by using photodetectors which can detect any emitted photon due to a decay of the form $L=\create{a}\hat{b}$.  Henceforth, we will use $\mathcal F^{(2)}_Q$ as the figure of merit. The objective of this work is to study the optimization of $\mathcal F^{(2)}_Q$ over initial states $\ket{\psi}$, for given $L$ and $p$. 

We consider a few standard examples. For a coherent state, $\mathcal F^{(2)}_Q(\ket{\psi_{\text{cohr}}}, S_z, L, p)= (1-p)N$. This can be interpreted as the standard quantum limit with the expected remaining number of atoms, after losses, which is given by $(1-p)N$.  For a GHZ state, it is straightforward to see that 
\begin{equation}
\mathcal F^{(2)}_Q(\ket{\psi_{\text{GHZ}}}, S_z, L, p) = (1-p)^N N^2
\end{equation}
for $L=\hat{a}, \hat{b}$ or $\create{a}\hat{b}$. Note the $N^2$ scaling, coming from the Heisenberg limit and the exponential scaling in $N$, coming from the fragility of the GHZ state.  One of the central challenges in quantum sensing is to develop protocols which retain the $N^2$ scaling (i.e., the Heisenberg scaling), but avoid the exponential scaling due to the loss. We show that there are other entangled states which are more robust to losses, in the sense that they avoid the exponential scaling in the loss while retaining a quantum advantage.

Two common entangled states used in quantum sensing are the  one axis twisting (OAT) and the two-axis counter twisting (TACT) states.  They are generated by evolving the coherent state under the OAT or the TACT Hamiltonians: $H_{OAT} = S_z^2$ and, $ H_{TACT} = S_y^2-S_z^2$.  Following ref.~\cite{PhysRevA.47.5138}, we define:
\begin{equation}
\begin{split}
\ket{OAT}& = e^{-i\theta_{oat}S_x}e^{-i\chi_{oat}S_z^2}\ket{\psi_{\text{cohr.}}} \text{ and } \\
\ket{TACT}&=e^{-i\theta_{tact}S_x}e^{-i\chi_{tact}(S_y^2-S_z^2)}\ket{\psi_{\text{cohr.}}}
\end{split}
\end{equation}
where, $\theta_{oat}, \chi_{oat}, \theta_{tact}$ and $\chi_{tact}$ are parameters optimized for $\mathcal F_Q$.  In the following section, we will derive general upper and lower bounds for $\mathcal F^{(2)}_Q$. 

 \section{Bounds on $\mathcal F^{(2)}_Q$}
As we mentioned before, there are two loss channels for a two component BEC: $L_a=\hat{a}$ and $L_2=\hat{b}$.  Therefore, the possible loss events in a given time interval are characterized by two integers $k_a, k_b$, representing the number of atoms lost through each channel. The probability of this event is given by 
\begin{equation}
P_{k_a, k_b}=\binom{N}{k_a}\binom{N}{k_b}p_a^{k_a}(1-p_a)^{N-k_a}p_b^{k_b}(1-p_b)^{N-k_b}
\end{equation}
This reduces to the previous case with a single loss channel when $p_b=0$. Moreover, Eq.~(\ref{F2}) generalizes to
\begin{equation}\label{F2_2}
\begin{split}
\mathcal F^{(2)}_Q(\ket{\psi}, S_z, \hat{a}, p_a, \hat{b}, p_b) &= \sum_{k_a, k_b=0}^N P_{k_a, k_b}\\
\times &\mathcal  F_Q \left(\frac{\hat{a}^{k_a}\hat{b}^{k_b}\ket{\psi}}{||\hat{a}^{k_a}\hat{b}^{k_b}\ket{\psi}||}, S_z\right)
\end{split}
\end{equation}
Note that Eq.~(\ref{F2}) and  Eq.~(\ref{F2_2}) are averaged over the two distributions shown in Fig.~\ref{Fig3}a. Intuitively, Eq.~(\ref{F2_2}) is averaged over a ``wider" distribution and is therefore expected to have a lower value.  We will consider these two cases, i.e., $p_a\approx p_b$ and $p_b\approx 0$ and show that $\mathcal F_Q^{(2)}$ is limited by the SQL scaling in the former, whereas in the latter case, it scales as $N^{3/2}$. We begin with the latter. 
 \subsection{The case $p_b\approx 0$} 
 This case corresponds to constraining the loss to $\hat{a}$.  The quantum Fisher information in any system is bound by the Heisenberg limit $\mathcal F_Q \leq N^2$ where $N$ is the number of atoms in the state.  This leads to an elementary upper bound:
 \begin{equation}\label{NoisyHL}
 \mathcal F^{(2)}_Q(\ket{\psi}, S_z, \hat{a}, p_a) \leq (1-p_a)^2N^2+(1-p_a)N-(1-p_a)^2N
 \end{equation}
 The R.H.S is obtained by summing $\sum_{k_a} \binom{N}{k_a}p_a^{k_a}(1-p_a)^{N-k_a}(N-k_a)^2$, since $(N-k_a)^2$ is the Heisenberg limit for the $k_a-$th term.  We refer to this limit as the \textit{noisy Heisenberg limit (Noisy HL)}.  Similarly, we define $(1-p_a)N$ as the noisy SQL. 
It is more important to show lower bounds for the maxima of $\mathcal F^{(2)}_Q(\ket{\psi}, S_z, \hat{a}, p_a)$, as they would shed light on the possible scaling with $N$.  We will show that 
\begin{equation}\label{lower_bound}
\max_{\ket{\psi}}\mathcal F^{(2)}_Q(\ket{\psi}, S_z, \hat{a}, p_a)\geq \frac{(1-p_a)^2 N^2}{1+\sqrt{p_a(1-p_a)N}}
\end{equation}
This result shows that it is possible to obtain a better-than-SQL scaling even in the presence of strong losses.  To show this bound, we need to develop some background. 

Note that the distribution $\{ \binom{N}{k_a}p_a^{k_a} (1-p_a)^{N-k_a} \}_{k_a}$ has an \textit{accumulation point} at $k_{a0}=\lfloor p_aN \rfloor$(Fig.~\ref{Fig3}a). That is, it has a peak at $k_{a0}$ with a waist of $\sqrt{p_a(1-p_a)N}$.  This means, in the sum in Eq.~(\ref{F2}), most of the contribution comes from around the peak, i.e., terms with $k_a \in \{k_{a0}-\sqrt{p_a(1-p_a)N}, \cdots, k_{a0}+\sqrt{p_a(1-p_a)N}\}$. Therefore, we can get a high $\mathcal F^{(2)}_Q$ if we find a state $\ket{\psi}$ such that the accumulation point $\ket{\psi_0}=\frac{\hat{a}^{k_{a0}}\ket{\psi}}{||\hat{a}^{k_{a0}}\ket{\psi}||}$ has a high quantum Fisher information.  We will use the notion of \textit{Moore-Penrose inverse } to determine such states.

$\ket{\psi_0}$ is an $N-k_{a0}$ atom state.  Moreover, the operator $\hat{a}$ is non-invertible and non-Hermitian.  In particular,  it maps an $N$ atom state to an $N-1$ atom state.  This means, the equation $\hat{a}\ket{\chi}=\ket{\psi_{0}}$ does not have a unique solution for $\ket{\chi}$.  However, we can pick a \textit{canonical solution} by minimizing the length of the vector \( \ket{\chi} \) among all solutions. This defines a pseudo-inverse to \( \hat{a} \) known as the Moore-Penrose inverse, which we denote by \( \hat{a}^{-MP} \) (the standard notation is \( \hat{a}^{+} \), which we avoid to prevent confusion with \( \hat{a}^{\dagger} \)). That is, \( \ket{\chi_0} = \hat{a}^{-MP} \ket{\psi_0} \) satisfies \( \hat{a} \ket{\chi_0} = \ket{\psi_0} \) while minimizing \( \langle \chi_0 \ket{\chi_0} \).  An equivalent definition of the Moore-Penrose inverse can be constructed using the singular value decomposition \( \hat{a} = U D V^{\dagger} \). Here, \( U \) and \( V \) are unitaries with potentially different dimensions, and \( D \) is a diagonal (possibly non-square) matrix with the singular values in the diagonal. The Moore-Penrose inverse can be defined as \( \hat{a}^{-MP} = V \tilde{D} U^{\dagger} \), where \( \tilde{D} \) is the diagonal matrix obtained by transposing \( D \) and replacing each non-zero diagonal entry with its inverse.   One can consider \( \hat{a}^{-MP} \) as the inverse of \( \hat{a} \) everywhere except in the kernel.

For a state $\ket{\psi_0}$ with $N-k_{a0}$ atoms,  we now define 
\begin{equation}
\ket{\psi} = \frac{(\hat{a}^{-MP})^{k_{a0}}\ket{\psi_0}}{||\hat{a}^{-MP})^{k_{a0}}\ket{\psi_0}||}
\end{equation}
as the canonical solution state.  Fig.~\ref{Fig3} shows the scaling of $\mathcal F^{(2)}_Q$ for the canonical solution state with $\ket{\psi_0}$ being the TACT state.  We will now prove Eq.~(\ref{lower_bound}) by showing that there exist states $\ket{\psi_0}$, such that the quantum Fisher information of $ \frac{(\hat{a}^{-MP})^{k_{a0}}\ket{\psi_0}}{||\hat{a}^{-MP})^{k_{a0}}\ket{\psi_0}||}$ is the R.H.S of Eq.~(\ref{lower_bound}).

Let $N_0=N-k_{a0}$ --- this is the mean number of atoms left.  We consider the GHZ state with $N0$ atoms, $\ket{\psi_{\text{GHZ}}}= \frac{(\create{b})^{N_0}}{\sqrt{2 N_0!}}\ket{\text{vec}}+\frac{(\create{a})^{N_0}}{\sqrt{2 N_0!}}\ket{\text{vec}}$. The quantum Fisher information is given by $\mathcal F(\ket{\psi_{\text{GHZ}}}, S_z)=N_0^2$.  The sum in Eq.~(\ref{F2}) is dominated by terms around $k_a=k_{a0}$,  in particular, terms with $|k_a-k_{a0}|\leq \sqrt{N_0p_a}$ make the dominant contribution. The corresponding to $k_a=k_{a0}-q$ is given by applying the Moore-Penrose inverse of $\hat{a}$ $q$ times
\begin{equation}
(\hat{a}^{-MP})^q\ket{\psi_{\text{GHZ}}} =  \frac{(\create{a})^q (\create{b})^{N_0}}{q! \sqrt{ N_0!}}\ket{\text{vac}}+\frac{(\create{a})^{N_0+q}}{ (N_0+q)!}\ket{\text{vac}}
\end{equation}
The quantum Fisher information is 
\begin{equation}
\mathcal F_Q((\hat{a}^{-MP})^q\ket{\psi_{\text{GHZ}}}, S_z) =4 N_0^2 \frac{\binom{N_0+q}{q}}{\left(1+\binom{N_0+q}{q}\right)^2} 
\end{equation}
Note that this scales as $N_0^2$ only when $q=0$.   It decreases with $q$ as $\sim \frac{1}{N_0^{q-2}}$. We can now sum these terms for $q=0, 1, \cdots, \sqrt{N_0p}$ with equal weights to find a lower bound for $\mathcal F^{(2)}_Q((\hat{a}^{-MP})^{k_{a0}}\ket{\psi_{\text{GHZ}}}, S_z, \hat{a}, p_a)$:
\begin{equation}
\begin{split}
\mathcal F^{(2)}_Q((\hat{a}^{-MP})^{k_{a0}}\ket{\psi_{\text{GHZ}}}, S_z, \hat{a}, p_a)& \geq\\
\frac{N_0^2}{ (1+\sqrt{N_0p_a})}& = \frac{(1-p_a)^2 N^2}{(1+\sqrt{N(1-p_a)p_a})}
\end{split}
\end{equation}
Note the $N^{3/2}$ scaling that appears asymptotically.  We will next show that $\mathcal F_Q^{(2)}$ is also bounded from above by $O(N^{3/2})$. That is, the above construction of the state using the Moore-Penrose inverse is also optimal.  More precisely, we will show that when $p_a>0$
\begin{equation}\label{upper_bound}
\lim_{N\rightarrow \infty} \frac{F^{(2)}_Q(\ket{\psi}, S_z, \hat{a}, p_a)}{N^{3/2}}<\infty
\end{equation}
holds for all states $\ket{\psi}$.  We present a heuristic sketch of the proof. Let us go back to the \textit{accumulation point}, i.e., $\ket{\psi_0}=\hat{a}^{k_{a0}}\ket{\psi}$ (we drop the normalization for simplicity of notation,  but we assume that all states are normalized).  $\mathcal F_Q^{(2)}$ gets most of the contribution from around the accumulation point.  Let us write this state in the standard basis $\ket{\psi_0}=\sum_n C_n \frac{(\create{a})^n (\create{b})^{N_0-n}}{\sqrt{n! (N_0-n)!}}\ket{\text{vac}}$ and define probabilities $\mu_n = |C_n|^2$.  For simplicity, we consider the operator $\hat{n}_a = \create{a}\hat{a}$ instead of $S_z$. Indeed, $\hat{n}_a = S_z + \frac{N_0}{2}$ and therefore, the quantum Fisher informations of $\hat{n}_a$ and $S_z$ are equal. That is, $\mathcal F_Q(\ket{\psi_0}, S_z)=\mathcal F_Q(\ket{\psi_0}, \hat{n}_a)$ and it can be written as  $\mathcal F_Q(\ket{\psi_0}, S_z) = 4 \left(\sum_n \mu_n n^2- \left(\sum_n \mu_n n\right)^2\right)$. Moreover after another loss event, $\mathcal F_Q(\hat{a}\ket{\psi_0}, S_z) = 4 \left(\frac{\sum_n \mu_n n^3}{\sum_n \mu_n n}-\left( \frac{\sum_n \mu_n n^2}{\sum_n \mu_n n}\right)^2\right)$. Note that all of $\mathcal F_Q(\hat{a}^{q}\ket{\psi_0}, S_z) $ can be expressed in terms of the moments of the distribution $\{\mu_n\}$, i.e., $M_q=\sum_n \mu_n n^q$, it follows that 
\begin{equation}
\mathcal F_Q(\hat{a}^q\ket{\psi_0}, S_z) \approx 4 \frac{M_{q+2}M_q -M_{q+1}^2}{M_q^2}
\end{equation}
We will now use a result regarding the scaling of the moments $M_q$ from probability theory (see refs. ~\cite{hardy1949divergent, wong1989asymptotic, lin2005tauberian}), which states that for generic distributions $\{\mu_n\}$, there exist constants $\lambda, \xi >0$ such 
\begin{equation}
M_q \leq \frac{\lambda}{q^{\xi}}N^q
\end{equation}
It follows now that the sum of the quantum Fisher informations is bounded by:
\begin{equation}\label{sum}
\sum_q \mathcal F_Q(\hat{a}^q\ket{\psi_0}, S_z)  \leq N^2 \times \mathcal O(1)\sum_q\frac{1}{q^2} = \mathcal O(N^2)
\end{equation}
The sum in Eq.~(\ref{F2}) contains about $\sqrt{Np_a(1-p_a)}$ significant terms and their sum is bounded by the above sum (Eq.~(\ref{sum})). Therefore, their \textit{average}, which is what we need for $\mathcal F^{(2)}_Q$ scales as $\mathcal O(N^2)/\sqrt{Np_a(1-p_a)} = \mathcal O(N^{3/2})$.

The intuitive picture is, in the sum Eq.~(\ref{F2}), there are $\mathcal O(\sqrt{N})$ significant terms around the accumulation point  (see Fig.~\ref{Fig3}a for a visual). Although the accumulation point may have a high quantum Fisher information (possibly $\mathcal F_Q \sim N^2$), the sum of the $\mathcal F_Q$s of these $\sqrt{N}$ terms can only be a constant factor higher than $N^2$, i.e., it can only be $\mathcal O(1)\times N^2$. This is because the quantum Fisher information declines rapidly as we go away from the accumulation point (mathematically, this is a consequence of the Hardy-Littlewood-Tauberian theorem~\cite{hardy1949divergent, wong1989asymptotic, lin2005tauberian}). Therefore, the average of these quantum Fisher informations can only scale as $\mathcal O(N^{3/2})$.  The same intuition leads to the no-go theorem when $p_a\approx p_b$, which we describe below.

%We define the uniform superposition state, $\ket{\psi_{\text{uni}}}=\frac{1}{\sqrt{N_0+1}}\sum_n \frac{(\create{a})^n (\create{b})^{N_0-n}}{\sqrt{n! (N_0-n)!}}\ket{\text{vac}}$, i.e., $C_n =\frac{1}{\sqrt{N_0+1}}$.  The leading order in the Fisher information corresponding to the state after $k$ loss events is 
%\begin{equation}
%\mathcal F(\ket{\psi_{\text{uni}}}, S_z) =  \frac{N_0^2}{12}+\frac{N_0}{6}
%\end{equation}
%The sum in Eq.~(\ref{F2}) is dominated by terms around $k=k_0$,  in particular, terms with $|k-k_0|\leq \sqrt{N_0p}$ make the dominant contribution.  After $k$ loss events, 
%\begin{equation}
%\mathcal F \left(\frac{\hat{a}^k \ket{\psi_{\text{uni}}}}{|| \hat{a}^k \ket{\psi_{\text{uni}}}||}, S_z\right) \approx N_0^2 \frac{k+1}{(k+3)(k+2)^2}.
%\end{equation}
%Note that while the scaling ($N_0^2$) is consistent with the Heisenberg limit,  it decreases with $k$ as $\sim \frac{1}{k^2}$. We can now sum these terms for $k=0, 1, \cdots, \sqrt{N_0p}$ with equal weights to find a lower bound for $\mathcal F^{(2)}(\ket{\psi_{\text{uni}}}, S_z, \hat{a}, p)$:
%\begin{equation}
%\begin{split}
%\mathcal F^{(2)}(\ket{\psi_{\text{uni}}}, S_z, \hat{a}, p)& \geq \\
%&\frac{1}{1+\sqrt{N_0p}}\sum_{k=0}^{\sqrt{N_0p}} N_0^2 \frac{k+1}{(k+3)(k+2)^2} \\
%\geq &\frac{N_0^2}{12 (1+\sqrt{N_0p})} = \frac{(1-p)^2 N^2}{12(1+\sqrt{N(1-p)p})}
%\end{split}
%\end{equation}
%\end{equation}

  \subsection{The case $p_b\approx p_a$}
This is a more general case and we have to evaluate the sum in Eq.~(\ref{F2_2}) over $k_a$ and $k_b$. The probability distribution $\{P_{k_a, k_b}\}$ is visually represented by the yellow surface in Fig.~\ref{Fig3}a. There is an accumulation point in this case as well, $(k_{b0}, k_{a0})=(\lfloor Np_b\rfloor, \lfloor Np_a\rfloor)$. However, the number of significant contributors around the accumulation point this time is $\sim \sqrt{Np_a(1-p_a)}\times \sqrt{Np_b(1-p_b)} = \mathcal O (N)$. This changes the upper limit on $\mathcal F_Q^{(2)}$ to $\mathcal O(N)$, thereby reducing it to at-most a  constant factor advantage over SQL.

 \section{Experimental considerations}
 
In this next section, we will briefly address the remaining two questions: (i) how do we prepare a Moore-Penrose inverted state? and (ii) how do we suppress the loss in one of the channels, i.e., get $p_b\approx 0$?  We begin with the former.

\subsection{Hamiltonian Engineering}
From the above proof of Eq.~(\ref{lower_bound}),  it is clear that for \textit{every} state $\ket{\psi}$ with $N_0=\lfloor (1-p_a) N\rfloor$ atoms and a quantum Fisher information that scales as $N_0^2$, there is an $N$ atom state constructed using the Moore-Penrose  inverse whose $\mathcal F^{(2)}_Q$ scales as $N^{3/2}$. More precisely, 
\begin{equation}
\mathcal F^{(2)}_Q\left(\frac{(\hat{a}^{-MP})^{k_{a0}}\ket{\psi}}{||\hat{a}^{-MP})^{k_{a0}}\ket{\psi}||}, S_z, \hat{a}, p_a)\right) \geq \frac{\mathcal F_Q(\ket{\psi}, S_z)}{1+\sqrt{N_0p_a}}.
\end{equation}
This gives us a wide range of $N$ atom states, all of which have an $\mathcal F^{(2)}_Q$ with a $3/2$ scaling, which can be useful in experimental implementations.  In this section we address the pertinent question how to experimentally prepare the Moore-Penrose inverse of a given state.  We demonstrate that elementary Hamiltonian engineering can be used to prepare most of such states~\cite{PhysRevLett.111.170502, PhysRevLett.114.240401, PhysRevX.14.031017, OKeeffe_2019}. We also numerically study the quantum Fisher information $\mathcal F^{(2)}_Q$ of the engineered states. 
\begin{figure}[h]
\includegraphics[scale=1.1]{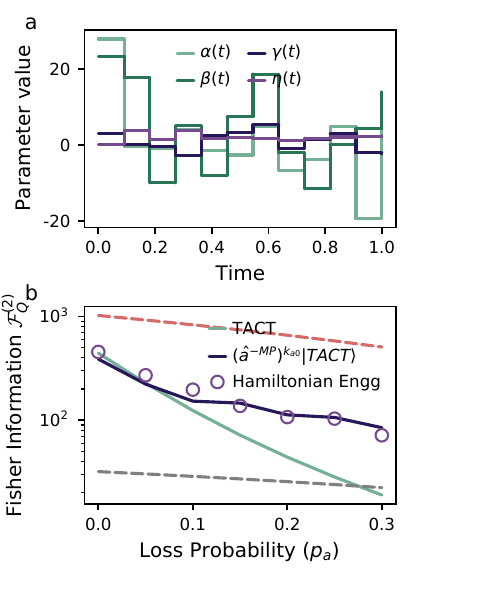}
\caption{\textbf{Hamiltonian engineering: }\textbf{a.} Optimized ${\alpha_i, \beta_i, \gamma_i, \eta_i}$ to prepare a Moore-Penrose inverse of the TACT state.  We use $m=10$ discrete points between $0, \tau$ in this example, achieving a typical fidelity of $0.98$. \textbf{b.} Quantum Fisher information $\mathcal F^{(2)}_Q$ for the engineered state compared to the TACT and the ideal Moore-Penrose inverse state.  The $F^{(2)}_Q$ of the Hamiltonian engineered state is very close to that of the Moore-Penrose inverse state --- a consequence of the high fidelity.  }\label{Fig4}
\end{figure}

It is well known that the operators $S_z=\frac{1}{2}(\create{a}\hat{a}-\create{b}\hat{b}),  S_x =  \create{a}\hat{b}+\create{b}\hat{a}$ and $S_y =-i(\create{a}\hat{b}-\create{b}\hat{a}) $ generate an $SU(2)$ subgroup within $SU(N+1)$.  It is also well known that the addition of a single quadratic term (e.g. $S_z^2, S_x^2, \{S_x, S_y\}$) completes the $SU(2)$ into the full group $SU(N+1)$. Therefore, any $SU(N+1)$ operator can be generated by evolution under the Hamiltonians $\{S_x, S_z, S_z^2\}$. The quadratic can be replaced by any other quadratic operator.  This is particularly useful, when we have to suppress the loss channel $\hat{b}$ by preventing t=interaction in that mode and using only $(\create{a}\hat{a})^2$ to generate the entanglement. 

For the purpose of demonstration, however,  we consider an \textit{over-complete} set, such as $\{S_x, S_z, S_z^2, \{S_x, S_z\}\}$ ($\{\cdot, \cdot\}$ is the anti-commutator). This bears an advantage in terms of faster convergence to the desired state.  We assume that the state preparation (see Fig.~\ref{Fig1}a) is done in time $\tau$ by applying the following Hamiltonian:
\begin{equation}
H(t) = \alpha(t) S_x + \beta(t) S_x + \gamma(t)S_z^2 +\eta(t) \{S_x, S_z\}
\end{equation}
where $\alpha, \beta$ and $\gamma$ are control functions to be optimized. We assume that in the experiment, these parameters can be changed only at discrete times. That is, we pick a sequence of times $0=t_1, t_2, \cdots, t_m=\tau$ regular interval (i.e., $t_{j+1}-t_j =\delta t$) and assume that the parameters $\alpha_j = \alpha(t_j)$, $\beta_j = \beta(t_j)$ and $\gamma_j = \gamma(t_j)$ etc can be controlled.  The unitary generated is $U = \Pi_j e^{-i\delta t (\alpha_j S_x + \beta_j S_z + \gamma_j S_z^2+\eta_j \{S_x, S_z\})}$. Therefore, the cost function is
\begin{equation}
\begin{split}
&f(\{\alpha_j, \beta_j, \gamma_j, \eta_j\}) = 1- \\
&\left| \frac{\bra{\psi}((\hat{a}^{-MP})^{k_{a0}})^{\dagger} | \Pi_j e^{-i\delta t (\alpha_j S_x + \beta_j S_z + \gamma_j S_z^2+\eta_j \{S_x, S_z\}))}\ket{\psi_{init}} }{||(\hat{a}^{-MP})^{k_{a0}} \ket{\psi} ||}\right|
\end{split}
\end{equation}
We set $\ket{\psi_{init}}=\ket{\psi_{\text{cohr.}}}$ and use the standard Broyden-Fletcher-Goldfarb-Shanno (BFGS) optimization to find the optimal $\{\alpha_j, \beta_j, \gamma_j, \eta_j\}$.   Fig. ~(\ref{Fig4}) shows the results for $\ket{\psi}=\ket{TACT}$.

\subsection{Constraining the atom loss}
There are two physical processes that cause loss of atoms in an atomic ensemble. (i) collisions of the atoms with background molecules and (ii) three-body loss~\cite{PhysRevA.69.023602}, i.e.,, loss of atoms due to strong interactions within the cloud.  The latter is more acute, especially when an entangling quantum control operation is applied during state preparation.  For instance, the squeezing Hamiltonians necessarily require the atoms to interact and therefore, this quantum control operation involves tuning some of the scattering lengths, which will invariable cause loss of atoms. 

We can constrain the loss to a single mode using a BEC double well --where the two wells form the two modes.  We can make one of the wells into a tighter trap. The atoms interact via s-wave scattering in the tighter trap and it can be used to entangle them. The s-wave scattering interaction also increases the collisional loss and therefore, any design to entangle the atoms will necessarily induce losses. However, by confining the loss to one of the two wells, we can retain a scalable quantum advantage in sensing using the techniques developed here. 

An alternative experimental implementation would be to use two spin components in a BEC and tune the Feshbach resonance to set the scattering length of collisions among one of the spin components to zero. This will still allow us to entangle the ensemble,  while constraining the loss to one mode. Again, the techniques developed here can be used to develop scalable sub-SQL magnetometers.

 \section{conclusions}
We have studied the problem of atom loss in quantum sensing with atomic ensembles and shown that if the loss is confined to a single component,  one can obtain a scalable quantum advantage with $\mathcal F_Q\sim N^{3/2}$, beating the SQL despite the losses. We have also developed a technique of constructing states that reach this scaling using a novel idea -- Moore Penrose inverse of the loss operator.   We have also shown that the $N^{3/2}$ scaling is indeed the upper bound for this loss,implying that our technique constructs the states with the optimal scaling. We also showed that the states that reach the optimal scaling can be prepared using Hamiltonian engineering. 

We also presented two experimental ideas to implement the confinement of atom loss to a single mode. --- one using a tunable double well potential and the other using Feshbach resonances. In either of these implementations, the suppression of interaction in one of the modes means that the atoms are allowed to interact only in the other mode.  While this leaves fewer choices for a control Hamiltonian, we have shown that the desired Moore-Penrose inverted states can be prepared using Hamiltonian engineering techniques, using interactions only in one spin mode.  Experimental implementation of the work presented here would be an exciting next step. In particular, it may open up the possibility of highly scalable quantum sensors. 

The technique of using a Moore-Penrose inverse can have applications in pre-mitigating other types of errors in quantum control. Recent developments in quantum error characterization~\cite{PhysRevResearch.6.043127} can be used to characterize the errors, which can then be mitigated using the Moore-Penrose inversion.

The idea of confining the loss to a single spin component hints towards a broader direction --- \textit{noise engineering}. This is another interesting future direction --- to explore the experimental techniques to minimize one type of noise at the cost of increasing another. Developing these techniques will ultimately pave the way for a deeper understanding of the effect of various types of noise on quantum information and can potentially lead to improved quantum advantage in various quantum technologies.

\section*{Acknowledgements}
We thank Ivan Deutsch,  Vikas Buchemmavari, Andrew Harter, Malcolm Boshier and Michael J. Martin for fruitful discussions.  Research presented in this article was supported by the Laboratory Directed Research and Development program of Los Alamos National Laboratory under project number 20230779PRD1.  Portions of this work were also supported by the U.S. Department of Energy, Office of Science, National Quantum Information Science Research Centers, Quantum Science Center (the computations related to Hamiltonian engineering). 

\paragraph*{\textbf{Data availability}} Data will be made available on request. 

\paragraph*{\textbf{Competing interests}} The authors declare no competing interests. 

\bibliography{References}

%\bibliography{references}

\end{document}